# Incremental Computation of Concept Diagrams

Francesco Kriegel

Theoretical Computer Science, TU Dresden, Germany

**Abstract.** Suppose a formal context $\mathbb{K} = (G, M, I)$ is given, whose concept lattice $\mathfrak{B}(\mathbb{K})$ with an attribute-additive concept diagram is already known, and an attribute column $\mathbb{C} = (G, \{n\}, J)$ shall be inserted to or removed from it. This paper introduces and proves an incremental update algorithm for both tasks.

**Keywords:** Formal Concept Analysis, Concept Diagram, Incremental Update, Column Insertion, Column Removal

## 1 Introduction

Every formal context $\mathbb{K} = (G, M, I)$ can be displayed by means of an (attribute-additive) diagram of its concept lattice $\mathfrak{B}(\mathbb{K})$. However, common algorithms focus on the computation of the concept set $\mathfrak{B}(\mathbb{K})$ or the concept neighborhood[1] $\prec$ as a whole, and do not provide any hints how to update the concept set, the concept neighborhood or even the concept diagram[2] upon changes in the underlying formal context.

Thus, each change would require a recomputation of the whole concept diagram. This means that unchanging fragments would be recomputed (which can be expensive), and furthermore it is then even not guaranteed that the unchanged parts of the concept diagram can be recognized as unchanged in the visualization by the user. To overcome this, I investigated the task of inserting or removing an attribute column into or from a formal context while updating the corresponding concept diagram with as little effort or visual changes as possible. The algorithm is called *iFox*,[3] and could further be used to deduce an update algorithm for setting or deleting just a single incidence entry in $\mathbb{K}$, or for adding or removing a bunch of attribute columns at once, or dualizing it to the insertion or removal of object rows. [4]

---

[1] The concept set may be ordered by extent inclusion, which yields a complete lattice $\underline{\mathfrak{B}}(\mathbb{K}) = (\mathfrak{B}(\mathbb{K}), \leq)$, see second section or [2] for further details. The concept neighborhood $\prec$ is the reflexive-transitive reduction of the concept order $\leq$.

[2] A concept diagram is a twice labeled directed acyclic graph $(\mathfrak{B}(\mathbb{K}), \prec, \gamma^{-1}, \mu^{-1})$ induced by the neighborhood relation on the concept set, together with a function that maps each node to a position into a vector space, and each node $(A, B)$ is labeled below by all objects $g \in G$, whose object concept $\gamma(g)$ equals $(A, B)$, and dually labeled above by all attributes $m \in M$ with $\mu(m) = (A, B)$.

[3] Historical note: In my time at SAP I implemented a FCA library called *fcaFox*, including an *iPred* algorithm. Thus, I chose the name *iFox* for my algorithm for the incremental computation of concept diagrams.

The next section gives some preliminaries on basic FCA and some lemmata for context appositions, the third section then formulates the necessary propositions to update the concept set, the neighborhood relation, the labels, the reducibility and seeds (for attributes, when drawing attribute-additive concept diagrams), and the arrow relations, respectivelly. Finally, the algorithm is formulated in pseudo code and its complexity is determined.

All lemmata and theorems can be found in, or are a condensed representation of, [4], except the last proposition describing the incremental update for the down arrows. The references further include some additional hints from the reviewers. This paper does not cover the incremental computation of pseudo-intents or implication bases. If you are interested in this topic, please have a look at [5].

## 2 Preliminaries

### 2.1 Basics of Formal Concept Analysis

A *formal context* $\mathbb{K} = (G, M, I)$ consists of two sets $G$ (*objects*) and $M$ (*attributes*), and furthermore a binary relation $I \subseteq G \times M$ (*incidence*) between them. For a pair $(g, m)$ that is enclosed in $I$, we also write $gIm$ and say that object $g$ *has* attribute $m$ (in context $\mathbb{K}$). A common visualization is a *cross table* as shown in the figure below on the left and another one on the right.

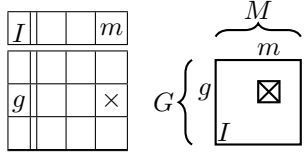

A *formal concept* $(A, B)$ of a context $\mathbb{K}$ consists of two sets, an *extent* $A \subseteq G$ and an *intent* $B \subseteq M$, such that their cartesian product $A \times B$ forms a maximal rectangle within the incidence relation $I$, more formally

$$A = B^I := \{g \in G \mid \forall_{m \in B}\, gIm\} \text{ and } B = A^I := \{m \in M \mid \forall_{g \in A}\, gIm\}.$$

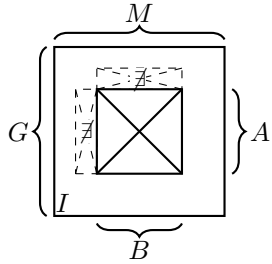

---

[4] If one wants to dualize the algorithm for row insertion or removal, and the concept diagram is still to be drawn attribute-additivelly, a characterization for the object reducibility update is neccessary. This can be found in [4].

The set of all formal concepts is denoted by $\mathfrak{B}(\mathbb{K})$, and this set can be ordered by means of the extents, i.e. concept $(A, B)$ is smaller than or equals concept $(C, D)$ iff extent $A$ is contained in extent $C$, symbol: $(A, B) \leq (C, D)$.

$\underline{\mathfrak{B}}(\mathbb{K}) := (\mathfrak{B}(\mathbb{K}), \leq)$ is a complete lattice and its infima and suprema are given by the equations

$$\bigwedge_{t \in T}(A_t, B_t) = \left(\bigcap_{t \in T} A_t, \left(\bigcup_{t \in T} B_t\right)^{II}\right) \text{ and } \bigvee_{t \in T}(A_t, B_t) = \left(\left(\bigcup_{t \in T} A_t\right)^{II}, \bigcap_{t \in T} B_t\right).$$

Sometimes the concept lattice of a given formal context shall be visualized for a highly structured and integrated view on its content. For this purpose the definition of a concept lattice is extended to the following notion of a concept diagram.

**Definition 1.** *Let $\mathbb{K}$ be a formal context and $\underline{V}$ a vector space, e.g. the real plane $\mathbb{R}^2$ or the real space $\mathbb{R}^3$ (or a discrete subset of them like $\mathbb{Z}^2$) for common visualizations. An* attribute-additive concept diagram *of $\mathbb{K}$ in $\underline{V}$ is a tuple*

$$\underline{\mathfrak{B}}_{\lambda, \sigma}(\mathbb{K}) := (\mathfrak{B}(\mathbb{K}), \prec, \lambda, \sigma)$$

*with the following components:*

1. *the concept lattice $(\mathfrak{B}(\mathbb{K}), \leq)$ and its neighborhood relation $\prec$,*
2. *the default label mapping (other choices possible, e.g. extent cardinality)*

$$\lambda \colon \begin{matrix} \mathfrak{B}(\mathbb{K}) \to \wp(G) \times \wp(M) \\ \mathfrak{b} \mapsto (\{\gamma = \mathfrak{b}\}, \{\mu = \mathfrak{b}\}) \end{matrix},$$

   *where all object labels in the first component $\gamma^{-1}(\mathfrak{b})$ are drawn below the concept $\mathfrak{b}$, and dually all attribute labels in the second component $\mu^{-1}(\mathfrak{b})$ are drawn above $\mathfrak{b}$,*
3. *and an arbitrary seed vector mapping $\sigma \colon M_{\mathrm{irr}} \to V$.*

*The position of a concept $(A, B)$ in $V$ is then defined as the sum of the seed vectors of all irreducible attributes in the intent, i.e.*

$$\pi(A, B) := \sum_{m \in B \cap M_{\mathrm{irr}}} \sigma(m).$$

### 2.2  Appositions of Formal Contexts

For two formal contexts $(G, M, I)$ and $(G, N, J)$ with disjoint attribute sets $M \cap N = \emptyset$ their *apposition* is defined as

$$(G, M, I) | (G, N, J) := (G, M \dot\cup N, I \dot\cup J).$$

**Lemma 2.** *Let $(G, M, I) | (G, N, J)$ be an apposition context, then the following equations hold for arbitrary objects $g \in G$ and attributes $m \in M$ and $n \in N$.*

1. $g(I \mathbin{\dot\cup} J)m \Leftrightarrow gIm$ and
   $g(I \mathbin{\dot\cup} J)n \Leftrightarrow gJn$
2. $g^{I \dot\cup J} = g^I \mathbin{\dot\cup} g^J$
3. $m^{I \dot\cup J} = m^I$ and
   $n^{I \dot\cup J} = n^J$

*Proof.* The proof is ommitted here, since the given equations are trivial.

**Lemma 3.** *Let $(G,M,I)|(G,N,J)$ be an apposition context and $A \subseteq G$ and $B \subseteq M \mathbin{\dot\cup} N$. Then the following equations hold:*

1. $A^{I \dot\cup J} \cap M = A^I$ and
   $A^{I \dot\cup J} \cap N = A^J$ and
   $A^{I \dot\cup J} = A^I \mathbin{\dot\cup} A^J$
2. $(B \cap M)^{I \dot\cup J} = (B \cap M)^I$ and
   $(B \cap N)^{I \dot\cup J} = (B \cap N)^J$ and
   $B^{I \dot\cup J} = (B \cap M)^I \cap (B \cap N)^J$
3. $A^{I(I \dot\cup J)} = (A^{I \dot\cup J} \cap M)^I = A^{II}$ and
   $A^{J(I \dot\cup J)} = (A^{I \dot\cup J} \cap N)^J = A^{JJ}$ and
   $A^{(I \dot\cup J)(I \dot\cup J)} = A^{II} \cap A^{JJ}$
4. $(B \cap M)^{I(I \dot\cup J)} = (B \cap M)^{II} \mathbin{\dot\cup} (B \cap M)^{IJ}$ and
   $(B \cap N)^{J(I \dot\cup J)} = (B \cap N)^{JI} \mathbin{\dot\cup} (B \cap N)^{JJ}$ and
   $B^{(I \dot\cup J)(I \dot\cup J)} = ((B \cap M)^I \cap (B \cap N)^J)^I \mathbin{\dot\cup} ((B \cap M)^I \cap (B \cap N)^J)^J$

*Proof.* The proof is obvious, use 2.

## 3   A Very Simple Example

Consider the free distributive lattice FCD(3) with three generating elements $x, y, z$, as shown in the figure below. An example is constructed that shows how an insertion and a removal of one attribute column affect the concept diagram.

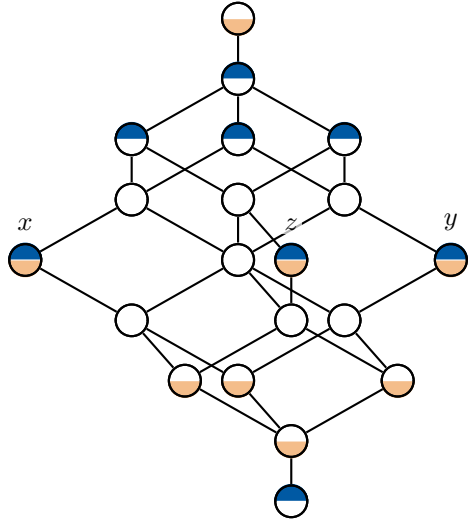

|  | $x \vee y \vee z$ | $x \vee y$ | $x \vee z$ | $y \vee z$ | $x$ | $y$ | $z$ | $\top$ |
|---|---|---|---|---|---|---|---|---|
| $x \wedge y \wedge z$ | × | × | × | × | × | × | × | ↗ |
| $y \wedge z$ | × | × | × | × | ↗ | × | × | |
| $x \wedge z$ | × | × | × | × | × | ↗ | × | |
| $x \wedge y$ | × | × | × | × | × | × | ↗ | |
| $z$ | × | ↗ | × | × |  |  | × | |
| $y$ | × | × | ↗ | × |  | × |  | |
| $x$ | × | × | × | ↗ | × |  |  | |
| $\top$ | ↗ |  |  |  |  |  |  | |

Choose all objects and the first six attributes as old context. The attribute $z$ is to be added. The appropriate contexts and their concept lattices are shown below.

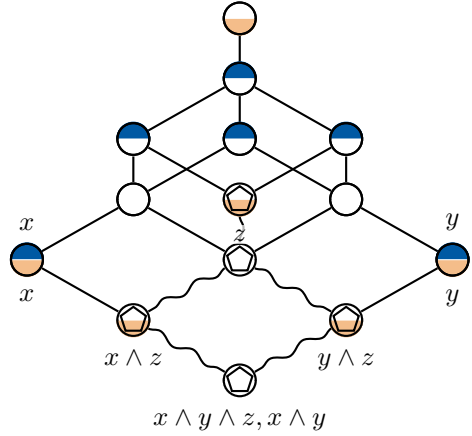

| $\mathbb{K}$ | $x \vee y \vee z$ | $x \vee y$ | $x \vee z$ | $y \vee z$ | $x$ | $y$ |
|---|---|---|---|---|---|---|
| $x \wedge y \wedge z$ | × | × | × | × | × | × |
| $y \wedge z$ | × | × | × | × | ↗ | × |
| $x \wedge z$ | × | × | × | × | × | ↗ |
| $x \wedge y$ | × | × | × | × | × | × |
| $z$ | × | ↗ | × | × | | |
| $y$ | × | × | ↗ | × | | × |
| $x$ | × | × | × | ↗ | × | |
| ⊤ | ↗ | | | | | |

In the initial state above some nodes are marked with a pentagon, these are the generator concepts. The final state below shows the concept lattice after insertion of column $z$, and the new concept nodes are marked with a star. As you can see the generator structure is locally doubled, and each new concept is a lower neighbor of its generator.

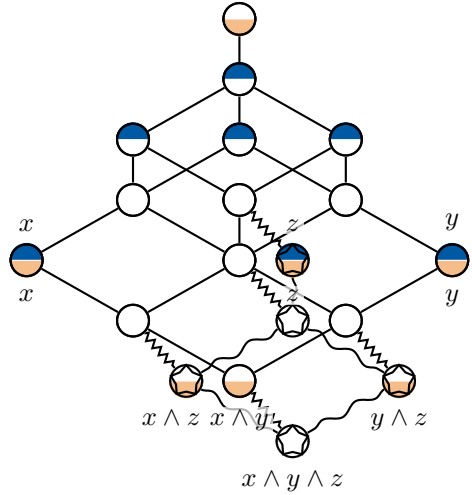

| $\mathbb{K}|\mathbb{C}$ | $x \vee y \vee z$ | $x \vee y$ | $x \vee z$ | $y \vee z$ | $x$ | $y$ | $z$ |
|---|---|---|---|---|---|---|---|
| $x \wedge y \wedge z$ | × | × | × | × | × | × | × |
| $y \wedge z$ | × | × | × | × | ↗ | × | × |
| $x \wedge z$ | × | × | × | × | × | ↗ | × |
| $x \wedge y$ | × | × | × | × | × | × | ↗ |
| $z$ | × | ↗ | × | × | | | × |
| $y$ | × | × | ↗ | × | | × | |
| $x$ | × | × | × | ↗ | × | | |
| ⊤ | ↗ | | | | | | |

## 4 Incremental Computation of Concept Diagrams

Throughout the whole section let $\mathbb{K} = (G, M, I)$ be an arbitrary formal context, called *old* context, with its concept diagram $(\mathfrak{B}(\mathbb{K}), \prec, \lambda, \sigma)$. Now the question arises what happens with the concept diagram when a new attribute column is inserted into $\mathbb{K}$, or when an existing attribute column is removed, respectivelly.

6      Francesco Kriegel

For this purpose let $n \notin M$ be the *new* attribute with its appropriate *column context* $\mathbb{C} = (G, \{n\}, J)$. The *new context* is then defined as the apposition $\mathbb{K}|\mathbb{C} := (G, M \,\dot\cup\, \{n\}, I \,\dot\cup\, J)$. [5] [6]

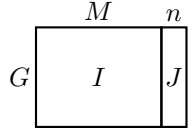

In the ongoing text we analyze the changes that occur on different levels of the concept diagram: concepts, neighborhood, labels, seeds, reducibility and arrows. Most of the main results are displayed in a table style: The old concept diagram on the left side and the new one on the right side, as shown below.

$$(\mathfrak{B}(\mathbb{K}), \prec, \lambda, \sigma) \quad \leftrightarrows \quad (\mathfrak{B}(\mathbb{K}|\mathbb{C}), \prec, \lambda, \sigma)$$

**Lemma 4.**  *1. For all object sets $A \subseteq G$ the following equivalence holds:*

$$A \subseteq n^J \Leftrightarrow A^J = \{n\}.$$

*2. For every concept $(A, B)$ of $\mathbb{K}|\mathbb{C}$ it holds that*

$$A \subseteq n^J \Leftrightarrow n \in B.$$

*Proof.*  1. Let $A \subseteq G$. Trivially $A^J \subseteq \{n\}$ always holds. The other set inclusion follows from the galois property, as $A \subseteq n^J$ is equivalent to $A^J \supseteq \{n\}$.
2. Let $(A, B)$ be an arbitrary concept of $\mathbb{K}$, i.e. $B = A^{I \dot\cup J} = A^I \,\dot\cup\, A^J$. Then by the first part, $A \subseteq n^J$ holds, iff $A^J = \{n\}$ holds. Obviously this implies $n \in B$. As $n \notin A^I$ always holds, $n \in B$ of course implies $A^J = \{n\}$.     □

### 4.1   Updating the Formal Concepts

First, we define a partition of the formal concept set of the old context $\mathbb{K}$, and dually a partition of the formal concept set of the new context $\mathbb{K}|\mathbb{C}$ and then formulate appropriate *update functions*, that map the parts of those partitions to each other. This then fully describes the update mechanism on the concept level from $\mathbb{K}$ to $\mathbb{K}|\mathbb{C}$ and vice versa.

**Definition 5.**  *A concept $(A, B)$ of $\mathbb{K}$ is called*

---

[5] For simplification of notion the set parenthesis of the singleton set $\{n\}$ may be omitted: The symbol $n$ is used both for the element $n$ itself and also for a singleton set containing this element $n$. It is always clear which variant is meant. We thus write $(G, n, J) := (G, \{n\}, J)$ for the column context, and $B \,\dot\cup\, n := B \cup \{n\}$ or else $B \setminus n := B \setminus \{n\}$ for an attribute set $B \subseteq M$.

[6] Sometimes both the old context $\mathbb{K}$ and the new context $\mathbb{K}|\mathbb{C}$ share the same set of concept extents; then $\mathbb{C}$ is called *redundant* für $\mathbb{K}$, and *irredundant* otherwise.

1. *old concept w.r.t.* $\mathbb{C}$, *iff its extent is no subset of the new attribute extent, i.e.* $A \nsubseteq n^J$,
2. *varying concept w.r.t.* $\mathbb{C}$, *iff* $A \subseteq n^J$, *and*
3. *generating concept w.r.t.* $\mathbb{C}$, *iff it is old and* $(A \cap n^J)^I = B$ *holds.*

The set of all old, varying and generating concepts is denoted by $\mathfrak{O}_\mathbb{C}(\mathbb{K})$, $\mathfrak{V}_\mathbb{C}(\mathbb{K})$ and $\mathfrak{G}_\mathbb{C}(\mathbb{K})$. Obviously every $\mathbb{K}$-concept is either old or varying, and each generating concept is particularly an old concept, i.e. $\{\mathfrak{O}_\mathbb{C}(\mathbb{K}), \mathfrak{V}_\mathbb{C}(\mathbb{K})\}$ is a partition of $\mathfrak{B}(\mathbb{K})$ and $\mathfrak{G}_\mathbb{C}(\mathbb{K}) \subseteq \mathfrak{O}_\mathbb{C}(\mathbb{K})$ holds.

**Definition 6.** *A concept* $(A, B)$ *of* $\mathbb{K}|\mathbb{C}$ *is called*

1. *old concept w.r.t.* $\mathbb{C}$, *iff its intent does not contain the new attribute, i.e.* $n \notin B$,
2. *varied concept w.r.t.* $\mathbb{C}$, *iff* $n \in B$ *and* $(B \setminus n)^I = A$, *and*
3. *generated (or new) concept w.r.t.* $\mathbb{C}$, *iff* $n \in B$ *and* $(B \setminus n)^I \neq A$.

The set of old, varied and generated concepts of $\mathbb{K}|\mathbb{C}$ is denoted by $\mathfrak{O}(\mathbb{K}|\mathbb{C})$, $\mathfrak{V}(\mathbb{K}|\mathbb{C})$ and $\mathfrak{G}(\mathbb{K}|\mathbb{C})$. We can easily see, that $\{\mathfrak{O}(\mathbb{K}|\mathbb{C}), \mathfrak{V}(\mathbb{K}|\mathbb{C}), \mathfrak{G}(\mathbb{K}|\mathbb{C})\}$ forms a partition of $\mathfrak{B}(\mathbb{K}|\mathbb{C})$.

As the names suggest, old concepts of $\mathbb{K}$ determine old concepts of $\mathbb{K}|\mathbb{C}$ and vice versa, $\mathbb{K}$-varying concepts determine $\mathbb{K}|\mathbb{C}$-varied concepts, and generating concepts from $\mathbb{K}$ induce new concepts of $\mathbb{K}|\mathbb{C}$. This is due to the following three bijections.

**Lemma 7.** *The following three mappings* $\mathfrak{o}$, $\mathfrak{g}$ *and* $\mathfrak{v}$ *are bijections.*

$$\mathfrak{B}(\mathbb{K}) \left\{ \begin{array}{c} \boxed{\mathfrak{O}_\mathbb{C}(\mathbb{K}) \quad A \nsubseteq n^J} \\ \subseteq \big\uparrow \\ \boxed{\mathfrak{G}_\mathbb{C}(\mathbb{K}) \quad \begin{array}{c} A \nsubseteq n^J \\ (A \cap n^J)^I = B \end{array}} \\ \\ \boxed{\mathfrak{V}_\mathbb{C}(\mathbb{K}) \quad A \subseteq n^J} \end{array} \begin{array}{c} (A,B) \mapsto (A,B) \\ \xrightarrow{\mathfrak{o}} \\ \xleftarrow{(A,B) \leftarrow (A,B)} \\ (A,B) \mapsto (A \cap n^J, B \,\dot\cup\, n) \\ \xrightarrow{\mathfrak{g}} \\ \xleftarrow{((B \setminus n)^I, B \setminus n) \leftarrow (A,B)} \\ (A,B) \mapsto (A, B \,\dot\cup\, n) \\ \xrightarrow{\mathfrak{v}} \\ \xleftarrow{(A, B \setminus n) \leftarrow (A,B)} \end{array} \begin{array}{c} \boxed{n \notin B \quad \mathfrak{O}(\mathbb{K}|\mathbb{C})} \\ \mathfrak{o} \circ \mathfrak{g} \big\uparrow \\ \boxed{\begin{array}{c} n \in B \\ (B \setminus n)^I \neq A \end{array} \mathfrak{G}(\mathbb{K}|\mathbb{C})} \\ \\ \boxed{\begin{array}{c} n \in B \\ (B \setminus n)^I = A \end{array} \mathfrak{V}(\mathbb{K}|\mathbb{C})} \end{array} \right\} \mathfrak{B}(\mathbb{K}|\mathbb{C})$$

*Proof.* Each of the following parts prove, that the mentioned mappings are well-defined and bijective. The original proof in [4] used the nested concept lattice of $\mathbb{C}$ in $\mathbb{K}$, the presented proof here is much simpler.

1. The mapping $\mathfrak{o}$ and its inverse are well-defined by lemma 4. The lower mapping is indeed the inverse, as we can easily see.

2. Let $(A, B)$ be a generating concept of $\mathbb{K}$ w.r.t. $\mathbb{C}$, then

$$(A \cap n^J)^{I \dot\cup J} = (A \cap n^J)^I \dot\cup (A \cap n^J)^J = B \dot\cup \{n\}$$

as surely $n \in (A \cap n^J)^J$ holds (because every object in $A \cap n^J$ has the new attribute $n$ w.r.t. $J$), and

$$(B \dot\cup \{n\})^{I \dot\cup J} = B^I \cap n^J = A \cap n^J.$$

Thus, the mapping $\mathfrak{g}$ is well-defined. The lower mapping is also well-defined by the following observation for an arbitrary generated concept $(A, B)$ of $\mathbb{K}|\mathbb{C}$, see also lemma 3

$$(B \setminus \{n\})^{II} = (B \cap M)^{II} = (A^{I \dot\cup J} \cap M)^{II} = A^{III} = A^I = \cdots = B \setminus \{n\}$$

Both mappings are inverse to each other, as can be seen on the intents.

3. Let $(A, B)$ be a varying concept of $\mathbb{K}$ w.r.t. $\mathbb{C}$, then for the extent we have $A^{I \dot\cup J} = A^I \dot\cup A^J = B \dot\cup \{n\}$ and for the intent we infer $(B \dot\cup \{n\})^{I \dot\cup J} = B^I \cap n^J = A \cap n^J = A$. Conversely for the lower mapping it holds that $A^I = A^{I \dot\cup J} \cap M = B \cap M = B \setminus \{n\}$ and $(B \setminus \{n\})^I = A$ by assumption. Both mappings are mutually inverse by looking on the extents. □

### 4.2 Updating the Neighborhood

Of course, when visualizing concept lattices, it is neccessary to update the concept neighborhood relation as well. Some first investigations show that there are blocks within the neighborhood that do not change from $\mathbb{K}$ to $\mathbb{K}|\mathbb{C}$ and vice versa. [7]

When inserting the new attribute, mainly the lower neighbors of the new concepts have to be computed. It is already clear that each new concept must be a lower neighbor of its generating concept. Also, each varied concept can not have any generator concept as upper neighbor.

For the attribute removal the columns and rows of new concepts of $\mathbb{K}|\mathbb{C}$ are just deleted, and the neighborhood between the varying and generator concepts needs to be determined.

A complete overview is given in the following figure (the bold subrelations change, and have to be computed; all other parts may be copied).

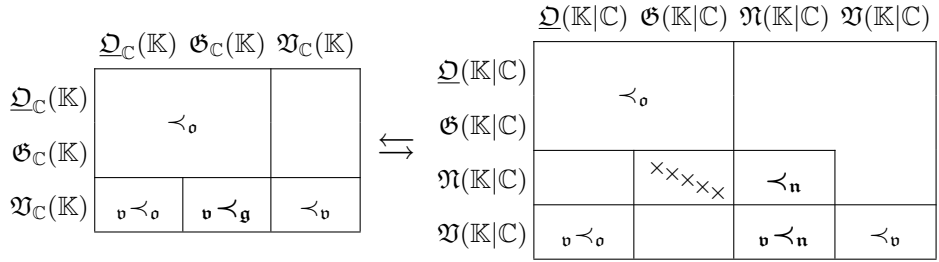

---

[7] It easy to see that the neighborhood between old concepts does not change, and so also for the varying/varied concepts.

Within the figure the *really old* concepts are used, that are just the old concepts which are no generator concept, denoted by

$$\underline{\mathfrak{O}}(\mathbb{K}|\mathbb{C}) := \mathfrak{O}(\mathbb{K}|\mathbb{C}) \setminus \mathfrak{G}(\mathbb{K}|\mathbb{C}) \quad \text{and} \quad \underline{\mathfrak{O}}_{\mathbb{C}}(\mathbb{K}) := \mathfrak{O}_{\mathbb{C}}(\mathbb{K}) \setminus \mathfrak{G}_{\mathbb{C}}(\mathbb{K}).$$

**Theorem 8.** *The concept neighborhood relation only changes partially:*

1. *Let $\mathfrak{a}, \mathfrak{b}$ be two generators in $\mathbb{K}$ w.r.t. $\mathbb{C}$, then $\mathfrak{n}(\mathfrak{a}) \prec_\mathfrak{n} \mathfrak{n}(\mathfrak{b})$ holds, iff*

$$[\mathfrak{a}, \mathfrak{b}] \cap \mathfrak{G}_{\mathbb{C}}(\mathbb{K}) = \{\mathfrak{a}, \mathfrak{b}\},$$

   *i.e. when there is no generating concept between $\mathfrak{a}$ and $\mathfrak{b}$.*
2. *If $\mathfrak{a}$ is varying and $\mathfrak{b}$ a generator, both in $\mathbb{K}$ w.r.t. $\mathbb{C}$, then $\mathfrak{v}(\mathfrak{a}) \,_\mathfrak{v}\!\prec_\mathfrak{n} \mathfrak{n}(\mathfrak{b})$ holds iff*

$$[\mathfrak{a}, \mathfrak{b}] \cap \mathfrak{G}_{\mathbb{C}}(\mathbb{K}) \cap \mathfrak{V}_{\mathbb{C}}(\mathbb{K}) = \{\mathfrak{a}, \mathfrak{b}\},$$

   *so if there is no generator or varying concept between $\mathfrak{a}$ and $\mathfrak{b}$.*
3. *Let $\mathfrak{a}$ be a varied concept and $\mathfrak{b}$ a new concept in $\mathbb{K}|\mathbb{C}$. Then $\mathfrak{v}^{-1}(\mathfrak{a}) \,_\mathfrak{v}\!\prec_\mathfrak{g} \mathfrak{g}(\mathfrak{b})$ holds in $\underline{\mathfrak{B}}(\mathbb{K})$ iff $\mathfrak{a} \,_\mathfrak{v}\!\prec_\mathfrak{n} \mathfrak{b}$ and*

$$(\mathfrak{a}, \mathfrak{og}(\mathfrak{b})) \cap \underline{\mathfrak{O}}(\mathbb{K}|\mathbb{C}) = \emptyset.$$

*Proof.* It is simply a proof by cases. The proof for the unchanging components is ommited here, and only the changing fragments are investigated. Some first clues can be obtained from the neighborhood structure within the nested concept lattice.

1. Let first $\mathfrak{a}$ and $\mathfrak{b}$ be two generating concepts. When are their generated new concepts neighboring? This can only be the case when no other concept is between them, and the only type of concept fitting between two new concepts is another new concept. In summary, the corresponding new concepts $\mathfrak{n}(\mathfrak{a})$ and $\mathfrak{n}(\mathfrak{b})$ are neighbors, iff there is no other generator concept between $\mathfrak{a}$ and $\mathfrak{b}$.
2. Analogously, let $\mathfrak{a}$ be a varying concept and $\mathfrak{b}$ a generating concept. Then the varied concept $\mathfrak{v}(\mathfrak{a})$ can only be covered by the new concept $\mathfrak{g}(\mathfrak{b})$, when there is no other $\mathbb{K}|\mathbb{C}$-concept between them. There could only be a varied or a new concept between them, and thus the statement holds exactly when there is no generator or varying concept between $\mathfrak{a}$ and $\mathfrak{b}$.
3. This is an immediate consequence of 2. For a varied concept $\mathfrak{a}$ and a new concept $\mathfrak{b}$, the corresponding varying concept $\mathfrak{v}^{-1}(\mathfrak{a})$ can only be covered by the generating concept $\mathfrak{g}(\mathfrak{b})$, when there is (in addition to the condition from 2) no really old concept between $\mathfrak{v}^{-1}(\mathfrak{a})$ and $\mathfrak{g}(\mathfrak{b})$, since this is the only missing concept type in the characterization of neighboring varied and new concepts, see 2. □

### 4.3   Updating the Labels

Each concept node is labeled with some objects and attributes. More exactly, each object concept $(g^{II}, g^{I})$ where $g \in G$ is labeled with $g$ above, and dually every attribute concept $(m^I, m^{II})$ where $m \in M$ is labeled with $m$ below.

When changing the context by column insertion or removal, the attribute label $n$ must be inserted in or removed from the concept diagram, and furthermore some other already existing labels might have to be moved to other concept nodes. In detail, the object concepts $\gamma(g)$ and the attribute concepts $\mu(m)$ have to be investigated to characterize the label update for the column insertion or removal. A complete overview for this is given in [4], and the condensed result is presented in the following proposition.

**Proposition 9.** *1. When adding the new attribute $n$, there must be an corresponding attribute concept $\mu(n)$ that is labeled with $n$. If $n$ is not redundant, then this new concept is always generated by the greatest generator concept*

$$\top_{\mathfrak{g}} := \bigvee \mathfrak{G}_{\mathbb{C}}(\mathbb{K}) = (n^{JII}, n^{JI}),$$

*and then $\mu(n) = \mathfrak{n}(\top_{\mathfrak{g}})$ holds.*
*2. For the concept diagram transition from $\mathbb{K}$ to $\mathbb{K}|\mathbb{C}$ only object labels at previously generator nodes can move downwards to the corresponding new concept node. No attribute labels change.*
*3. Vice versa, for the transition from $\mathbb{K}|\mathbb{C}$ back to $\mathbb{K}$ the attribute label $n$ is removed and the object labels of a generator concept are merged with the object labels of the approriate new concept, i.e. let $(A, B)$ be a generator with object labels $L$ and $(C, D)$ the generated new concept with object labels $M$, then $(A, B)$ is labeled with each element from the union $L \cup M$ in the old concept diagram.*

|  | G M |  | $n^{J\mathbb{C}}$ | $n^J$ | M | n |
|---|---|---|---|---|---|---|
| $\mathfrak{D}_{\mathbb{C}}(\mathbb{K})$ | $\lambda_{\mathfrak{o}}$ | $\mathfrak{O}(\mathbb{K}|\mathbb{C})$ |  | $\lambda_{\mathfrak{o}}$ |  |  |
| $\mathfrak{G}_{\mathbb{C}}(\mathbb{K})$ | $\lambda_{\mathfrak{g}}$ $\longleftrightarrow$ | $\mathfrak{G}(\mathbb{K}|\mathbb{C})$ | $\lambda_{\mathfrak{g}}$ |  | $\lambda_{\mathfrak{g}}$ |  |
| $\mathfrak{V}_{\mathbb{C}}(\mathbb{K})$ | $\lambda_{\mathfrak{v}}$ | $\mathfrak{N}(\mathbb{K}|\mathbb{C})$ |  | $\lambda_{\mathfrak{g}}$ |  | $\times$ $\top_{\mathfrak{g}}$ |
|  |  | $\mathfrak{V}(\mathbb{K}|\mathbb{C})$ |  | $\lambda_{\mathfrak{v}}$ |  |  |

*Proof.* This is easy and straight-forward by analyzing the object and attribute concepts, and determining whether they are old, varying/varied or generating/new.                                                                                              □

### 4.4   Updating the Reducibility and Seeds

In order to maximize the quality of an attribute-additive concept diagram it is important to know the irreducible attributes of the context. Each attribute can then be displayed as the infimum of irreducible attributes, and thus, the set of irreducible attributes spans the whole concept diagram and it suffices to assign seed vectors just to the irreducible attributes. Of course, when inserting or removing $\mathbb{C}$ to $\mathbb{K}$ or from $\mathbb{K}|\mathbb{C}$, the attribute irreducibility may change for the existing attributes.

**Proposition 10.** *The attribute reducibility can be updated via the following observations:*

1. *Each $\mathbb{K}$-reducible attribute is also $\mathbb{K}|\mathbb{C}$-reducible.*
2. *A $\mathbb{K}$-irreducible attribute $m \in M$ is $\mathbb{K}|\mathbb{C}$-reducible, iff its $\mathbb{K}$-attribute concept is varying and the corresponding unique upper neighbor $\mu_\mathbb{K}^*(m)$ is really old, and furthermore at least one superconcept of $\mu_\mathbb{K}^*(m)$ is a generator concept.*
3. *Every $\mathbb{K}|\mathbb{C}$-irreducible attribute is also $\mathbb{K}$-irreducible.*
4. *A $\mathbb{K}|\mathbb{C}$-reducible attribute $m \in M$ is $\mathbb{K}$-irreducible, iff its $\mathbb{K}|\mathbb{C}$-attribute concept is varied and has exactly one old upper neighbor $\mathfrak{b}$ and overthis only new upper neighbors, that are generated from superconcepts of $\mathfrak{b}$.*

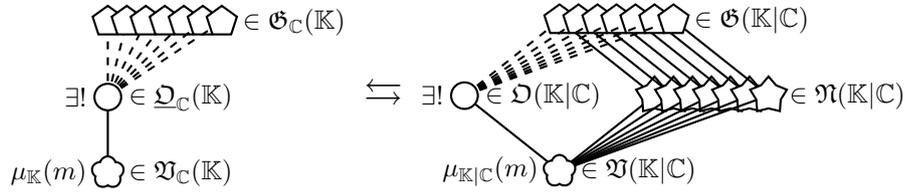

*Proof.* 1. First, if $m$ is a $\mathbb{K}$-reducible attribute, then the attribute extent $m^I$ can be obtained by an intersection of attribute extents $\bigcap_{m \in B} m^I$ with $m \notin B$. Obviously then also

$$m^{(I \dot\cup J)} = m^I = \bigcap_{m \in B} m^I = \bigcap_{m \in B} m^{(I \dot\cup J)}$$

holds, hence $m$ is $\mathbb{K}|\mathbb{C}$-reducible.

2. Second, let $m$ be a $\mathbb{K}$-irreducible attribute.

($\Rightarrow$) Suppose $m$ is $\mathbb{K}|\mathbb{C}$-reducible. If $\mu_\mathbb{K}(m)$ were an old concept, then $\mu_{\mathbb{K}|\mathbb{C}}(m) = \mathfrak{o}(\mu_\mathbb{K}(m))$ and the set of upper neighbors does not change according to theorem 8. Thus, the irreducibility of $m$ in $\mathbb{K}$ implies the irreducibility of $m$ in $\mathbb{K}|\mathbb{C}$. Contradiction! Hence, the attribute concept $\mu_\mathbb{K}(m)$ must be varying. By 7, there are no other old or varied upper neighbors of $\mu_{\mathbb{K}|\mathbb{C}}(m)$. If $\mu_\mathbb{K}^*(m)$ would be a varying or generating concept, then

$$\mu_{\mathbb{K}|\mathbb{C}}(m) = \mathfrak{v}(\mu_\mathbb{K}(m)) \prec \begin{cases} \mathfrak{v}(\mu_\mathbb{K}^*(m)) & \text{if } \mu_\mathbb{K}^*(m) \in \mathfrak{V}_\mathbb{C}(\mathbb{K}) \\ \mathfrak{g}(\mu_\mathbb{K}^*(m)) & \text{if } \mu_\mathbb{K}^*(m) \in \mathfrak{G}_\mathbb{C}(\mathbb{K}) \end{cases}$$

holds. Let $\mathfrak{b} \in \mathfrak{G}_\mathbb{C}(\mathbb{K})$ with $\mathfrak{b} \neq \mu_\mathbb{K}^*(m)$, such that $\mathfrak{g}(\mathfrak{b})$ covers $\mu_{\mathbb{K}|\mathbb{C}}(m)$, then $\mu_\mathbb{K}(m)$ must be a lower neighbor of $\mathfrak{b}$ and there is no varying or generating concept between them. So $\mu_\mathbb{K}(m) \prec \mu_\mathbb{K}^*(m) < \mathfrak{b}$ must hold, but this is a contradiction. In summary, $\mathfrak{v}(\mu_\mathbb{K}^*(m))$ or $\mathfrak{g}(\mu_\mathbb{K}^*(m))$, respectivelly, must be the unique upper neighbor of $\mu_{\mathbb{K}|\mathbb{C}}(m)$, and $m$ would be $\mathbb{K}|\mathbb{C}$-irreducible. Contradiction! Hence $\mu_{\mathbb{K}}^*(m)$ must be an old non-generator concept. Finally if there were no generating superconcept above $\mu_{\mathbb{K}}^*(m)$, then $\mathfrak{o}(\mu_{\mathbb{K}}^*(m))$ were the only upper neighbor of $\mu_{\mathbb{K}|\mathbb{C}}(m)$, *i.e.* $m$ would be $\mathbb{K}|\mathbb{C}$-irreducible. Contradiction!

($\Leftarrow$) Suppose the attribute concept $\mu_{\mathbb{K}}(m)$ is a varying concept and its unique upper neighbor $\mu_{\mathbb{K}}^*(m)$ is an old non-generator concept that has at least one generator superconcept. Denote the minimal ones of these generator superconcepts by $\xi_1, \xi_2, \ldots, \xi_k$. Then the following structure on the left side can be found within the concept lattice of $\mathbb{K}$. Neighboring concept nodes are connected by straight line segments and comparable concepts are connected by zig zag line segments. Then according to theorem 8 the new concepts $\mathfrak{g}(\xi_1), \ldots, \mathfrak{g}(\xi_k)$ must cover the varied attribute concept $\mathfrak{v}(\mu_{\mathbb{K}}(m))$. This is due to the fact, that no varying concept can be greater than an old concept, and the generators $\xi_1, \ldots, \xi_k$ are minimal. Furthermore $\mu_{\mathbb{K}}^*(m)$ is the unique upper neighbor of $\mu_{\mathbb{K}}(m)$, hence there cannot be any varying or generating concept between $\mu_{\mathbb{K}}(m)$ and each $\xi_j$. In summary, the transition from $\mathbb{K}$ to $\mathbb{K}|\mathbb{C}$ changes the concept lattice structure as displayed in the right diagram. Obviously $\mu_{\mathbb{K}|\mathbb{C}}(m) = \mathfrak{v}(\mu_{\mathbb{K}}(m))$ has more than one upper neighbor, hence $m$ is $\mathbb{K}|\mathbb{C}$-reducible.

3. Let first $m \in M$ be a $\mathbb{K}|\mathbb{C}$-irreducible attribute. Then $m$ must also be $\mathbb{K}$-irreducible, as otherwise $m$ were $\mathbb{K}|\mathbb{C}$-irreducible by 1.

4. Second, let $m \in M$ be $\mathbb{K}|\mathbb{C}$-reducible attribute.

($\Rightarrow$) Suppose $m$ is $\mathbb{K}$-irreducible. Then $\mu_{\mathbb{K}|\mathbb{C}}(m)$ must be a varied concept. Otherwise $\mu_{\mathbb{K}}(m) = \mathfrak{o}^{-1}(\mu_{\mathbb{K}|\mathbb{C}}(m))$ were an old concept and this is a contradiction to 1. If $\mu_{\mathbb{K}|\mathbb{C}}(m)$ had more than one old (and thus non-generating) upper neighbor in $\mathfrak{B}(\mathbb{K}|\mathbb{C})$, then the according old concepts in $\mathfrak{B}(\mathbb{K})$ would cover $\mu_{\mathbb{K}}(m)$. This is a contradiction to the $\mathbb{K}$-irreducibility of $m$. So $\mu_{\mathbb{K}|\mathbb{C}}(m)$ has exactly one old upper neighbor $\omega \in \mathfrak{O}(\mathbb{K}|\mathbb{C})$, all other upper neighbors must be varied or new concepts. If a varied concept covers $\mu_{\mathbb{K}|\mathbb{C}}(m)$, then its appropriate varying concept covers $\mu_{\mathbb{K}}(m)$ as well. Again, this is a contradiction to the $\mathbb{K}$-irreducibility. So all other upper neighbors must be new concepts. If there were any new concept $\nu \in \mathfrak{G}(\mathbb{K}|\mathbb{C})$ whose generator $\xi$ is not a superconcept of $\omega$, then $\mu_{\mathbb{K}}(m)$ would be covered by $\mathfrak{o}^{-1}(\xi)$. Then $\mu_{\mathbb{K}}(m)$ had at least two upper neighbors and this contradicts the $\mathbb{K}$-irreducibility.

($\Leftarrow$) Suppose $\mu_{\mathbb{K}|\mathbb{C}}(m)$ varies and has exactly one upper neighbor $\omega$ and over-this only new upper neighbors $\nu_1, \ldots, \nu_k$, whose generators are greater than $\omega$. Then choose $\xi_j := \mathfrak{g}(\nu_j)$ and the same structure as in the right diagram above occurs, and by theorem 8 $\mathfrak{o}^{-1}(\omega) = \mu_{\mathbb{K}}^*(m)$ must be the unique upper neighbor of $\mu_{\mathbb{K}}(m)$. This means $m$ is $\mathbb{K}$-irreducible. □

The update of the seed map can now be done with the following rules.

1. When adding the new column, delete the seeds for $\mathbb{K}|\mathbb{C}$-reducible attributes, that were $\mathbb{K}$-irreducible, and introduce a new seed for $n$.

2. When removing the column, delete the seed for $n$ and compute seeds for the previously reducible attributes in $\mathbb{K}|\mathbb{C}$, which are now irreducible in $\mathbb{K}$.

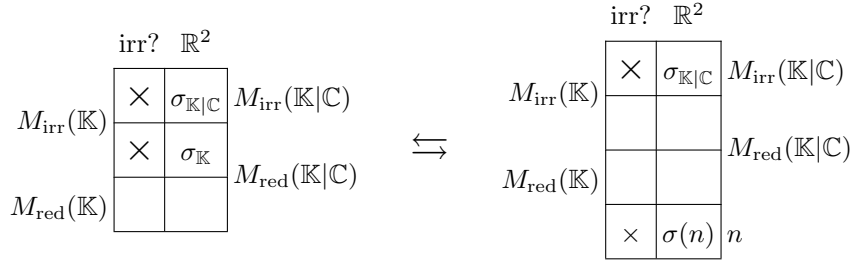

## 5  Incremental Computation of the Arrow Relations

### 5.1  Updating the Up Arrows

This section investigates the changes for the up arrow relation. For this purpose the object set and the attribute set is splitted into the following components:

$$G_1 := \{g \mid g \notin n^J\},\ G_2 := \{g \mid g \in n^J\},\ \text{and}$$
$$M_1 := \{m \mid m^I \not\subset n^J\},\ M_2 := \{m \mid m^I \subset n^J\}$$

When the column is inserted the block $\nearrow_{\mathbb{K}} \subseteq G_1 \times M_2$ can simply be deleted. The only entries to compute is the upper column $\nearrow_n \subseteq G_1 \times \{n\}$. [8] It is even possible to give a characterization for the $\nearrow_{\mathbb{K}}$ block for the column removal.

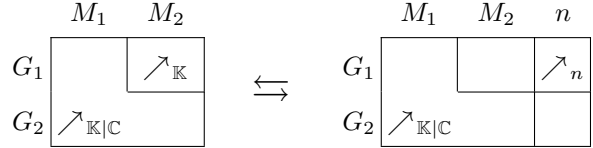

**Proposition 11.**  *1. Up arrows in $\mathbb{K}$ and $\mathbb{K}|\mathbb{C}$ may only differ on the subset $G_1 \times M_2$ and $G_1 \times \{n\}$. All other parts are equal.*
*2. Let $g \in G_1$ and $m \in M_2$, then $g \nearrow_{\mathbb{K}} m$ holds, iff one of the following conditions is fulfilled:*
  *(a) $m$ is $\mathbb{K}|\mathbb{C}$-reducible, and its attribute concept $\mu_{\mathbb{K}|\mathbb{C}}(m) \in \mathfrak{V}(\mathbb{K}|\mathbb{C})$ has exactly one old upper neighbor $\mathfrak{b}$ and overthis only new upper neighbors generated by superconcepts of $\mathfrak{b}$, and furthermore $\gamma_{\mathbb{K}|\mathbb{C}}(g)$ is a subconcept of $\mathfrak{b}$.*

---
[8] Of course, there cannot be any arrows in the lower column $G_2 \times \{n\}$ as it is full of crosses.

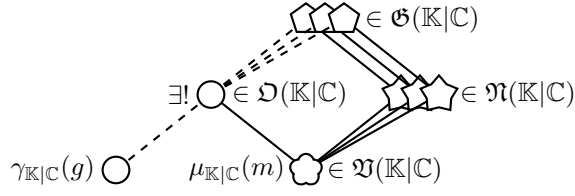

(b) $m$ is $\mathbb{K}|\mathbb{C}$-irreducible, $\mu_{\mathbb{K}|\mathbb{C}}^*(m) \in \mathfrak{N}(\mathbb{K}|\mathbb{C})$ and the old object concept $\gamma_{\mathbb{K}|\mathbb{C}}(g) \in \mathfrak{O}(\mathbb{K}|\mathbb{C})$ is a subconcept of the generator $\mathfrak{og}(\mu_{\mathbb{K}|\mathbb{C}}^*(m))$.

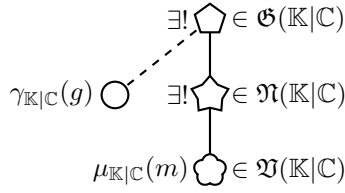

*Proof.* 1. This is obvious.
2. In case $g \in n^J$ this follows from the preceding lemma as well. Suppose $g \notin n^J$. Then the object concept of $g$ in $\mathbb{K}|\mathbb{C}$ is given by

$$\gamma_{\mathbb{K}|\mathbb{C}}(g) = \begin{cases} \mathfrak{o}(\gamma_{\mathbb{K}}(g)) & \text{if } \gamma_{\mathbb{K}}(g) \in \mathfrak{O}_{\mathbb{C}}(\mathbb{K}) \\ \mathfrak{v}(\gamma_{\mathbb{K}}(g)) & \text{if } \gamma_{\mathbb{K}}(g) \in \mathfrak{V}_{\mathbb{C}}(\mathbb{K}) \end{cases}.$$

(a) Let $m$ be $\mathbb{K}|\mathbb{C}$-reducible. $g \nearrow_{\mathbb{K}} m$ can only hold, when $m$ is irreducible in $\mathbb{K}$, i.e. when $\mu_{\mathbb{K}|\mathbb{C}}(m) \in \mathfrak{V}(\mathbb{K}|\mathbb{C})$ has exactly one old upper neighbor $\omega$ and overthis only new upper neighbors, whose generators are superconcepts of $\omega$, according to 10. Then $\mathfrak{o}^{-1}(\omega)$ is the unique upper neighbor of $\mu_{\mathbb{K}}(m)$. Furthermore, $\gamma_{\mathbb{K}|\mathbb{C}}(g) \leq \omega$ holds, iff $\gamma_{\mathbb{K}}(g) \leq \mu_{\mathbb{K}}^*(m)$, i.e. iff $g \nearrow_{\mathbb{K}} m$.
(b) When $m$ is $\mathbb{K}|\mathbb{C}$-irreducible, then $m$ is also $\mathbb{K}$-irreducible by 10. Furthermore, $g \notin n^J$ implies $g \not\nearrow_{\mathbb{K}|\mathbb{C}} m$, i.e. $\gamma_{\mathbb{K}|\mathbb{C}}(g)$ is no subconcept of $\mu_{\mathbb{K}|\mathbb{C}}^*(m)$. If $\mu_{\mathbb{K}|\mathbb{C}}^*(m)$ is an old concept, then $\mathfrak{o}^{-1}(\mu_{\mathbb{K}|\mathbb{C}}^*(m))$ is the unique upper neighbor of

$$\mu_{\mathbb{K}}(m) = \begin{cases} \mathfrak{o}^{-1}(\mu_{\mathbb{K}|\mathbb{C}}(m)) & \text{if } \mu_{\mathbb{K}|\mathbb{C}}(m) \in \mathfrak{O}(\mathbb{K}|\mathbb{C}) \\ \mathfrak{v}^{-1}(\mu_{\mathbb{K}|\mathbb{C}}(m)) & \text{if } \mu_{\mathbb{K}|\mathbb{C}}(m) \in \mathfrak{V}(\mathbb{K}|\mathbb{C}) \end{cases}.$$

Then $\gamma_{\mathbb{K}}(g)$ is a subconcept of $\mu_{\mathbb{K}}^*(m)$, iff $\gamma_{\mathbb{K}|\mathbb{C}}(g)$ is a subconcept of $\mu_{\mathbb{K}|\mathbb{C}}^*(m)$. As this cannot occur according to the preconditions, $g \not\nearrow_{\mathbb{K}} m$ must hold. If $\mu_{\mathbb{K}|\mathbb{C}}^*(m)$ is a varied concept, then $\mathfrak{v}^{-1}(\mu_{\mathbb{K}|\mathbb{C}}^*(m))$ is the unique upper neighbor of $\mu_{\mathbb{K}}(m) = \mathfrak{v}^{-1}(\mu_{\mathbb{K}|\mathbb{C}}(m))$. Then $\gamma_{\mathbb{K}}(g)$ is smaller than $\mu_{\mathbb{K}}^*(m)$, iff $\gamma_{\mathbb{K}|\mathbb{C}}(g)$ is a subconcept of $\mu_{\mathbb{K}|\mathbb{C}}^*(m)$. Thus, $g \not\nearrow_{\mathbb{K}} m$ as well in this case. If the unique upper neighbor $\mu_{\mathbb{K}|\mathbb{C}}^*(m)$ is a new concept, then according to 8 $\mathfrak{g}(\mu_{\mathbb{K}|\mathbb{C}}^*(m))$ must be the unique upper neighbor of $\mu_{\mathbb{K}}(m) = \mathfrak{v}^{-1}(\mu_{\mathbb{K}|\mathbb{C}}(m))$. Furthermore $\gamma_{\mathbb{K}}(g)$ can only be a subconcept of $\mu_{\mathbb{K}}^*(m)$, if it is an old concept and a subconcept of the generator. (If $\gamma_{\mathbb{K}}(g)$ would be varying and smaller than the generator, $\gamma_{\mathbb{K}|\mathbb{C}}(g)$ must be smaller than the new generated concept as well, in contradiction to the preconditions.) In summary, $g \nearrow_{\mathbb{K}} m$ holds in this case, iff $\gamma_{\mathbb{K}|\mathbb{C}}(g)$ is an old concept and smaller than the generator of the upper neighbor of $\mu_{\mathbb{K}|\mathbb{C}}(m)$. □

### 5.2 Updating the Down Arrows

Suppose, $g \in G$ is an object and $m \in M$ is an attribute of $\mathbb{K}$. First, observe that by definition of the down arrows it holds that

$$g \swarrow_{\mathbb{K}} m \Leftrightarrow g \not I m \text{ and } \bigvee_{h \in G} g^I \subsetneq h^I \Rightarrow h\, I\, m$$

and analogously

$$g \swarrow_{\mathbb{K}|\mathbb{C}} m \Leftrightarrow \underbrace{(g,m) \notin (I \dot\cup J)}_{\Leftrightarrow g \not I m} \text{ and } \bigvee_{h \in G} g^{I \dot\cup J} \subsetneq h^{I \dot\cup J} \Rightarrow \underbrace{h\, (I \dot\cup J)\, m}_{h I m}.$$

**Proposition 12.**  *1. When $g \swarrow_{\mathbb{K}|\mathbb{C}} m$ holds, then also $g \swarrow_{\mathbb{K}} m$ holds.*
*2. Let $g \swarrow_{\mathbb{K}} m$ where $g \not J n$. Then $g \swarrow_{\mathbb{K}|\mathbb{C}} m$ holds, if there is no $\mathbb{K}$-equivalent object $h$ (i.e. $g^I = h^I$), which is not $\mathbb{K}|\mathbb{C}$-equivalent to $g$ (i.e. $h\, J\, n$).*
*3. Let $g \swarrow_{\mathbb{K}} m$ where $g\, J\, n$. Then $g \swarrow_{\mathbb{K}|\mathbb{C}} m$ holds, if each object $h \in G$ with $g^I \subsetneq h^I$ also has the new attribute $n$.*

*Proof.* 1. This is obvious, since $g^I \subsetneq h^I$ implies $g^{(I \dot\cup J)} \subsetneq h^{(I \dot\cup J)}$.
 2. Suppose $g$ does not have the new attribute $n$, and $g \swarrow_{\mathbb{K}} m$ holds. When does $g \swarrow_{\mathbb{K}|\mathbb{C}} m$ also hold? For $h \in G$ with $g^{I \dot\cup J} \subsetneq h^{I \dot\cup J}$ it holds that $g^I \subsetneq h^I \dot\cup h^J$.
   – If $g^I \subsetneq h^I$, then $h\, I\, m$ holds since $g \swarrow_{\mathbb{K}} m$.
   – If $g^I = h^I$ and $h\, J\, n$, then $h \not I m$ since $g$ does not have $m$ (as $g \swarrow_{\mathbb{K}} m$ holds).
   Obviously $g \swarrow_{\mathbb{K}|\mathbb{C}} m$ cannot hold, when the second condition is fulfilled.
 3. Finally, let $g$ have the new attribute $n$ and $g \swarrow_{\mathbb{K}} m$. To check, whether $g \swarrow_{\mathbb{K}|\mathbb{C}} m$ hold, let $h \in G$ be an object, whose $\mathbb{K}|\mathbb{C}$-intent is a proper superset of $g^{I \dot\cup J}$. It then easily follows, that also $h$ must have the new attribute $n$ and $g^I \subsetneq h^I$ must hold for the $\mathbb{K}$-intents. By the precondition this yields $h\, I\, m$. Since this is true for all such objects $h$, $g \swarrow_{\mathbb{K}|\mathbb{C}} m$ can be concluded. □

## 6 Conclusion

This document described an update algorithm for the insertion or removal of an attribute column to or from a formal context, whose concept diagram is already known. It has been implemented in *ConceptExplorer FX*, that is a partial re-implementation of the well-known FCA tool *ConceptExplorer* by Serhiy Yevtushenko et al.

The introduced lemmata and propositions may be extended for the insertion or removal of several attribute columns at once, or it may be dualized for object row insertion or deletion, as also suggested in the introduction. Furthermore it may be possible to generalize it to insert elements into an arbitrary complete lattice, not only to insert new attribute concepts into a concept lattice (and also for deletion, of course).